\documentclass{article}
\usepackage{dcase2019,amsmath,graphicx,url,times,booktabs, tabularx}
\def\UrlBreaks{\do\/\do-}
\usepackage{subcaption,placeins}

\title{MULTI-SOURCE DIRECTION-OF-ARRIVAL ESTIMATION USING IMPROVED ESTIMATION CONSISTENCY METHOD}

\name{Rohith Mars$^{1}$,
      Hiroyuki Ehara$^{2}$,
       Srikanth Nagisetty$^{1}$, 
       Chong Soon Lim$^{1}$,
       }
\address{$^1$ Panasonic R\&D Center, Singapore \{rohith.mars, srikanth.nagisetty, chongsoon.lim\}@sg.panasonic.com\\          
         $^2$ Panasonic Corporation, Japan \enspace ehara.hiroyuki@jp.panasonic.com\\ 
  }

\begin{document}

\ninept
\maketitle

\begin{sloppy}

\begin{abstract}
We address the problem of estimating direction-of-arrivals (DOAs) for multiple acoustic sources in a reverberant environment using a spherical microphone array. It is well-known that multi-source DOA estimation is challenging in the presence of room reverberation, environmental noise and overlapping sources. In this work, we introduce multiple schemes to improve the robustness of estimation consistency (EC) approach in reverberant and noisy conditions through redefined and modified parametric weights. Simulation results show that our proposed methods achieve superior performance compared to the existing EC approach, especially when the sources are spatially close in a reverberant environment.
\end{abstract}


\section{Introduction}
\label{sec:intro}

One of the long-standing and most challenging research topic in acoustic signal processing is the estimation of direction-of-arrival (DOA) of multiple source signals using microphone arrays. The estimated DOA of the acoustic sources in a soundfield has applications in robotics and surveillance, where they can be used for acoustic source tracking, source separation, dereverberation, beamforming and speech enhancement~\cite{benesty2008microphone,brandstein2013microphone}. Under practical conditions, the DOA estimation of multiple acoustic sources is challenging. This can be attributed to the effects of room reverberation and reflections, presence of background and microphone additive (self) noise. In addition, the estimation is made more challenging in the presence of simultaneously active or overlapped in acoustic activity of the sources as well as in conditions where the multiple sources are spatially close to each other. In literature, most of the proposed techniques for source localization employ uniformly-spaced omni-directional microphone arrays arranged in a linear, circular or planar configuration. However, in the past decade there has been a growing interest in acoustic source localization using spherical microphone arrays (SMAs), where the microphones are mounted on a rigid baffle of spherical configuration~\cite{rafaely2015fundamentals,jarrett2017theory}. Compared to conventional microphone arrays, SMAs possess the ability to fully capture a soundfield in three dimensions and represent the captured soundfield in terms of higher order ambisonics (HOA) for further processing and rendering. 

Generally, the captured SMA signals in time-domain are first converted to the frequency domain (e.g., using FFT) and then transformed and further processed using the spherical harmonic transform (SHT) to generate the spherical harmonic coefficients (SHC) or HOA signals. The DOA of the acoustic sources are then estimated using the computed HOA signals. Among the various previously proposed approaches of DOA estimation using SMAs, the pseudo-intensity vector (PIV) method has been the most attractive in terms of computational complexity as well as localization accuracy. The PIV is an estimate of the active intensity vector and is computed from the HOA signals. Denoting the PIV as $\mathbf{I}$, the DOA unit vector can be computed as $\mathbf{u} = \mathbf{I}/||\mathbf{I}||$, where $\mathbf{u}$ is the unit vector and $||\cdot||$ represents the vector norm. For more information on spherical harmonics and estimation of PIV, the reader is suggested to refer to~\cite{jarrett2017theory, williams1999fourier,jarrett20103d}.

In~\cite{evers2014multiple}, the PIVs are estimated for all the time-frequency (TF) points in the signal spectrum. To detect source direction of multiple sources, the estimated PIVs (and correspondingly the DOA unit vectors) are clustered (for example, using K-means clustering). The estimated cluster centroids are then treated as the direction of the sources. However, this approach has limited performance under the presence of room reverberation and background noise. In~\cite{nadiri2014localization,pavlidi20153d}, a similar approach is used. In addition to the PIV-based approach, it includes techniques such as direct path dominance (DPD) test, coherence test and histogram smoothing to improve the accuracy of DOA estimation under room reverberation and background noise. However, the coherence test and smoothing are computationally expensive as it involves computation of second-order statistics of the signal. In addition, the smoothening of histogram requires prior knowledge about the source spacing as well. It was also reported that the performance is limited when the acoustic sources are closely spaced. To address the above complexity as well as performance limitations, a low-complex alternative based on estimation consistency (EC) approach~\cite{hafezi2017multiple} which involves post-processing on the DOA unit vectors estimated for each TF bin is used for DOA estimation. This post-processing includes the estimation of parameters to identify the DOA unit vectors corresponding to single source-dominant TF points.

In this work, we propose techniques to further improve the performance of EC approach. More specifically, we redefine and modify the weighting parameters that were used in the EC approach so that they become more robust under the conditions of reverberation and additive noise. We propose three schemes to improve the EC approach and evaluate each of them in terms of DOA estimation accuracy of multiple acoustic sources using simulations performed using an SMA in a reverberant environment.

This paper is organized as follows. Section~\ref{EC} briefly reviews the EC-based approach for DOA estimation. In Section~\ref{sec:proposed}, we discuss the limitations of the EC-based approach and propose three schemes to improve the performance of the same. The details of the simulations and the evaluation results are discussed in Section~\ref{sec:sim_results} with the conclusions provided in Section~\ref{conclusion}.

\section{Review of Estimation consistency approach}
\label{EC}
Estimation consistency (EC) approach~\cite{hafezi2017multiple} is based on post-processing of the  DOA unit vectors estimated at each TF point. The post-processing involves estimation of multiple parameters to discriminate between single source-dominant and multi-source/noise-dominant time-frames as well as identification of TF points with more accurate DOA information. 

As the initial step, for each time-frame $\tau$, an estimation is performed to identify whether the time-frame consists of a single source or multiple sources (including noise). This is performed by computing the coefficient-of-variation parameter~\cite{ahonen2009diffuseness}. In~\cite{hafezi2017multiple}, this parameter is chosen as the average of the DOA unit vectors in a particular time-frame. The average DOA unit vector for a given time-frame,  $\Hat{\mathbf{u}}(\tau)$ is computed as
\begin{equation}
    \Hat{\mathbf{u}}(\tau) = \dfrac{1}{K}\sum_{k=1}^{K} \mathbf{u}(k,\tau),
\end{equation}
\noindent where $k= 1,\hdots, K$ is the frequency bin index.
The norm of $\Hat{\mathbf{u}}(\tau)$ is then used to estimate whether the time-frame consists of a single source or not. If $||\Hat{\mathbf{u}}(\tau)||\approx 1$, it suggests that most of the DOA unit vectors in time-frame $\tau$ correspond to a specific direction and hence this time-frame can be considered as a single source-dominant frame. When $||\Hat{\mathbf{u}}(\tau)||\approx 0$ it suggests that the DOA unit vectors are pointing to random directions due to the presence of noise/multiple sources, thereby minimizing the norm of the average DOA unit vector. Using $\Hat{\mathbf{u}}(\tau)$, a weighting $\psi(\tau)$ is computed to identify the time-frames dominated by single-source as
\begin{equation}
    \psi(\tau) = 1-\sqrt{1-||\Hat{\mathbf{u}}(\tau)||}.
    \label{eq:psi}
\end{equation}
The value of $\psi(\tau)$ denotes if the frame consists of a single source or multiple sources/noise.  From~(\ref{eq:psi}), it is easy to note that $\psi(\tau)$ close to 1 denotes the presence of single source, while $\psi(\tau)$ close to 0 signifies the presence of multiple sources or a noise-dominant frame.

\par As the next step, an additional weighting is used to identify the frequency bins which contribute more accurately to the DOA estimation in a given time-frame. To achieve this, the angular deviation of the DOA unit vectors (for multiple frequency bins) from the average DOA unit vector corresponding to each time-frame is computed. A higher weighting is applied if the angular deviation is small and vice-versa. Denoting this within-frame weighting as $\lambda(k,\tau)$, it is computed as
\begin{equation}
    \lambda(k,\tau) = 1-\frac{1}{\pi}\cos^{-1}\biggr(\dfrac{\mathbf{u}(k,\tau)^{T}\Hat{\mathbf{u}}(\tau)}{||\mathbf{u}(k,\tau)||||\Hat{\mathbf{u}}(\tau)||}\biggr).
\end{equation}
The final EC weights $ w_\mathrm{{EC}}(k,\tau)$ were then estimated as the product of $\psi(\tau)$ and $\lambda(k,\tau)$ as
\begin{equation}
    w_\mathrm{{EC}}(k,\tau) = \psi(\tau) \times  \lambda(k,\tau).
\end{equation}

\noindent In summary, a higher $w_\mathrm{{EC}}(k,\tau)$ will be given to TF points belonging to a time-frame with a dominant single source (decided by $\psi(\tau)$) and closest to the average DOA (decided by $\lambda(k,\tau)$) in the time-frame. 

The estimation of $w_\mathrm{{EC}}(k,\tau)$ is performed as described above for each TF point. Based on these EC weights, a subsampling is subsequently performed to select the DOA unit vectors with accurate DOA information. More specifically, the DOA unit vectors with the highest $P\%$ weight is chosen from the entire set of DOA unit vectors. In~\cite{hafezi2017multiple}, the value of $P$ was chosen empirically. In the final step, the subsampled DOA unit vectors, are  clustered (assuming that the number of sources are known \emph{apriori}) to estimate the cluster centroids. Each cluster centroid is then treated as the DOA unit vector and is associated to the corresponding source.

\section{Proposed Method}
\label{sec:proposed}
\begin{figure*}%
\centering
\begin{subfigure}{\columnwidth}
\includegraphics[width=\columnwidth]{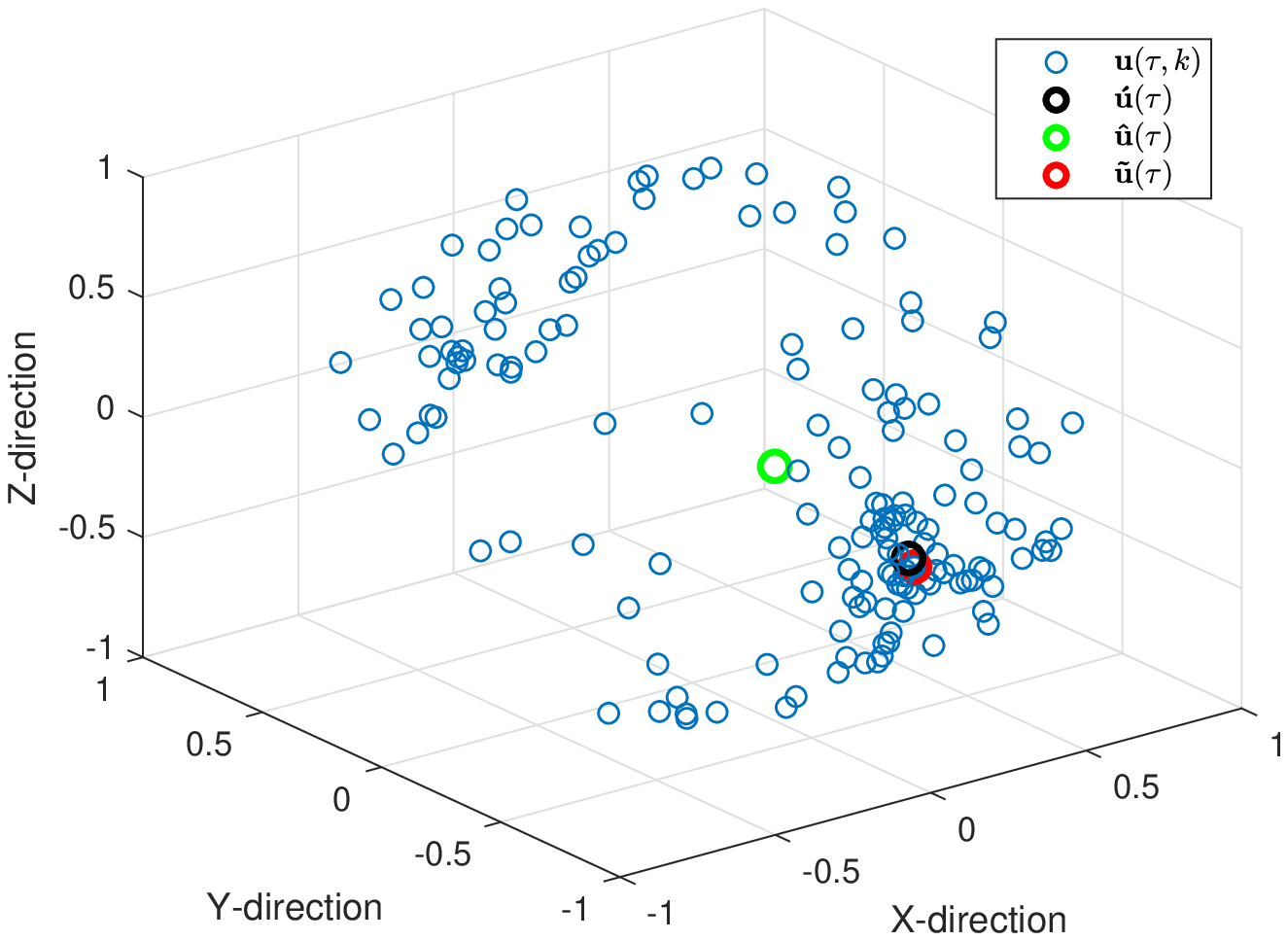}%
\caption{}%
\label{subfiga}%
\end{subfigure}\hfill%
\begin{subfigure}{\columnwidth}
\includegraphics[width=\columnwidth]{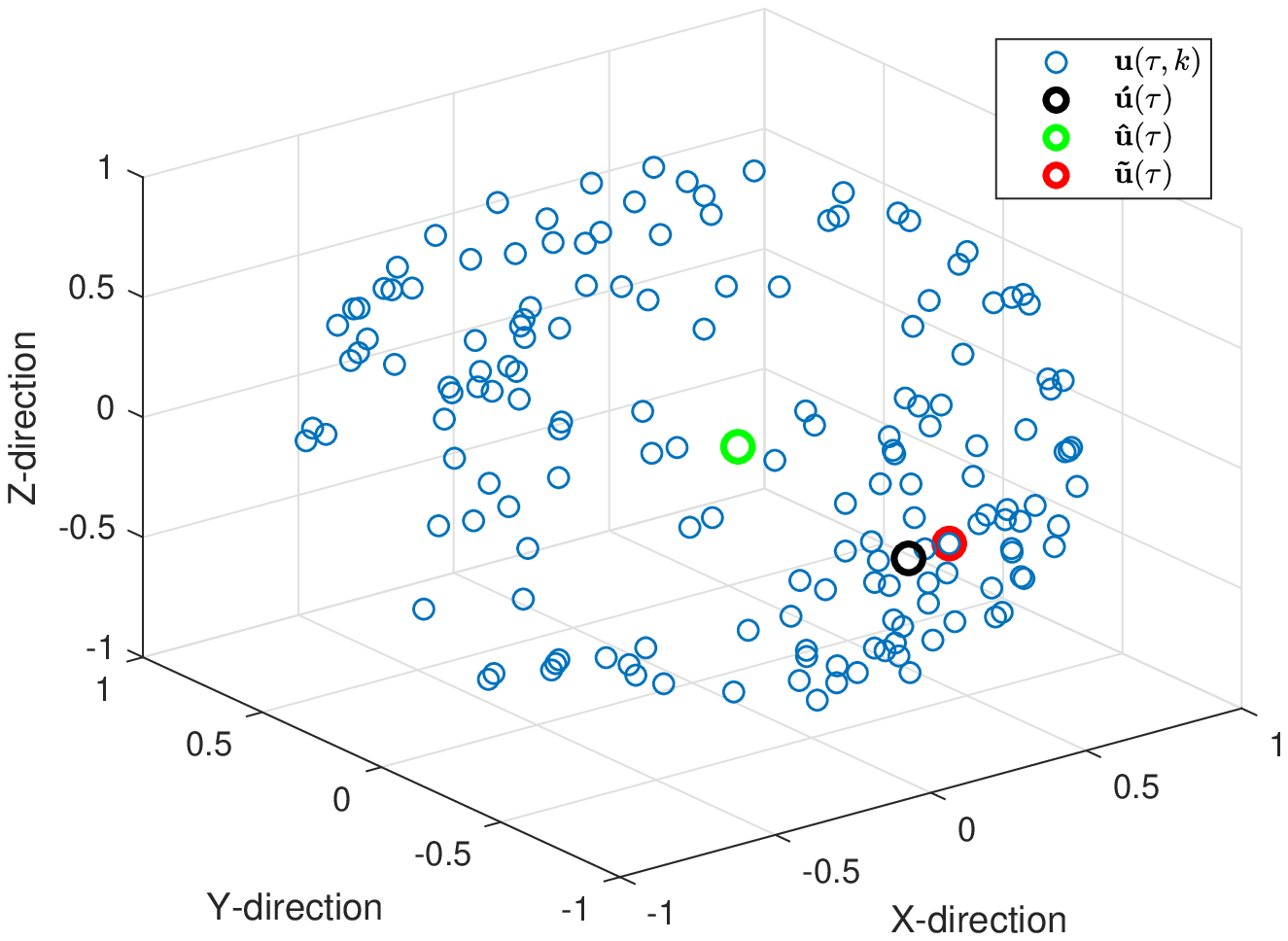}%
\caption{}%
\label{subfigb}%
\end{subfigure}\hfill%
\caption{Scatter plot showing the distribution of DOA unit vectors  $\mathbf{u}(k,\tau)$ along with the ground truth DOA unit vector $ \Acute{\mathbf{u}}(\tau)$, average DOA unit vector $ \Hat{\mathbf{u}}(\tau)$ and DOA unit vector belonging to a cluster of accurate DOA points $\Tilde{\mathbf{u}}(\tau)$ for reverberation time fixed at (a) $200$ ms (b) $400$ ms.}
\label{fig:effect_reverb}
\end{figure*}
In our proposed method, we introduce three schemes aimed to improve the performance of EC approach in closed environments with spatially close sources. More specifically, we redefine the EC weighting parameter $\lambda(\tau,k)$ and introduce modifications on $\psi(\tau)$ such  that the resulting EC weights are more robust to such conditions. 

\par In~\cite{hafezi2017multiple}, the average DOA unit vector $ \Hat{\mathbf{u}}(\tau)$ was used to compute the EC weights. However, it is well-known that statistical average is sensitive to outliers. Under practical conditions of noise and/or reverberation, the DOA unit vectors will be pointing at random directions. In addition, for the case of speech signals, there will be signal sparsity, i.e., the speech signal will be active only for certain frequency bins. In such conditions, the use of  $\Hat{\mathbf{u}}(\tau)$  will lead to a bias (from the ground truth DOA unit vector) depending on the amount of erroneous DOA unit vectors. 

To validate the above hypothesis, we performed multi-condition simulations using a spherical microphone array (SMA) placed in a room consisting of a single speech source with room reverberation time fixed at (a) $200~\mathrm{ms}$ and (b) $400~\mathrm{ms}$. We fix the location of the speech source so that it can be represented by the ground truth DOA unit vector $\Acute{\mathbf{u}}(\tau)$. The use of single speech source ensures that the variation in the estimated DOA unit vectors are caused only by the presence of reverberation and speech sparsity and not due to the presence of multiple sources. For both conditions, we manually chose a time-frame where speech signal is active and then observed the variation in DOA unit vectors at each TF point in this time-frame using scatter plots as shown in Figure~\ref{fig:effect_reverb}. From the scatter plots we make the following observations - (1) the presence of reverberation and signal sparsity results in DOA unit vectors $\mathbf{u}(k,\tau)$ to point in random directions, (2) the average DOA unit vector $ \Hat{\mathbf{u}}(\tau)$ is biased from the ground truth DOA unit vector $\Acute{\mathbf{u}}(\tau)$ (3) a few DOA unit vectors point close to the ground truth direction, forming a cluster which corresponds to more accurate DOA unit vectors.

From the above observations, we note that a more robust estimate of $\Acute{\mathbf{u}}(\tau)$ must belong in the cluster of accurate DOA unit vectors. The identification of this DOA unit vector belonging to the cluster $ \Tilde{\mathbf{u}}(\tau)$ is achieved by utilizing the L2-distance (Euclidean distance) criterion as
\begin{equation}
     \Tilde{\mathbf{u}}(\tau) = \mathbf{u}( k_i,\tau)  ~\text{where}~i = \min_{i} (\sum_{j=1}^{K}||\mathbf{u}( k_i,\tau) - \mathbf{u}( k_j,\tau)||.
\end{equation}
\noindent From the same scatter plot in Figure~\ref{fig:effect_reverb}, it can be seen that for both conditions,  $\Tilde{\mathbf{u}}(\tau)$ is a better estimate of the ground truth and is a more robust estimate as compared to  $\Hat{\mathbf{u}}(\tau)$ under the simulated reverberant conditions. 
\par In the following subsections, we propose three  schemes which utilize the above defined $\Tilde{\mathbf{u}}(\tau)$ and further modification on $\psi(\tau)$ to improve the EC-weights $w_\mathrm{{EC}}(k,\tau)$.

\subsection{Improved EC-1}
\par In this scheme, we use $\psi(\tau)$ computed using the average DOA unit vector to identify the time-frames dominated by single source. However, we use the above obtained  $\Tilde{\mathbf{u}}(\tau)$ to estimate the redefined within-frame weighting $ \Bar{\lambda}(k,\tau)$ as

\begin{equation}
    \Bar{\lambda}(k,\tau) = 1-\frac{1}{\pi}\cos^{-1}\biggr(\dfrac{\mathbf{u}(k,\tau)^{T}\Tilde{\mathbf{u}}(\tau)}{||\mathbf{u}(k,\tau)||||\Tilde{\mathbf{u}}(\tau)||}\biggr).
\end{equation}
The use of $\Tilde{\mathbf{u}}(\tau)$ ensures that the DOA unit vectors closer to the cluster of accurate DOA unit vectors are given higher weighting and vice-versa.
\noindent The improved EC weights, hereafter termed as $w_{\mathrm{EC-1}}(k,\tau)$ are then estimated as the product of $\psi(\tau)$ and $\Bar{\lambda}(k,\tau)$ as

\begin{equation}
    w_{\mathrm{EC-1}}(k,\tau) = \psi(\tau) \times \Bar{\lambda}(k,\tau).
\end{equation}

\subsection{Improved EC-2}
In this scheme, the parameter  $\Bar{\lambda}(k,\tau)$ is estimated similar to EC-1 approach. However, additional processing is performed on the $\psi(\tau)$ parameter. In order to further improve the weighting, clustering is performed on $\psi(\tau)$ to identify time-frames with more probability of being dominated by a single source. We use K-means clustering on the estimated $\psi(\tau)$ values. By setting the number of clusters to two, it is possible to obtain two cluster centroids $\psi_0(\tau)$ and $\psi_1(\tau)$ where $\psi_1(\tau)>\psi_0(\tau)$. Time-frames with more dominance of a single source can be identified as time-frames for which the corresponding $\psi(\tau)$ is greater than $\psi_1(\tau)$.  These time-frames are then assigned a new weighting of $1$ and those time-frames frames with $\psi(\tau)$ less than $\psi_1(\tau)$ are assigned a weighting of $0$ to form binary weights $\Bar{\psi}(\tau)$. A schematic of the estimation of $\Bar{\psi}(\tau)$ after the clustering of $\psi(\tau)$ is shown in Figure~\ref{fig:kmeans}. It can be seen that for time-frames with higher probability of being single-source dominant is given a boost in the weighting, while the weights corresponding to the other time-frames are suppressed.

Since the clustering technique is used only to find the cluster centroids, K-means clustering can also be replaced by Fuzzy c-means clustering. The cluster centroids estimated from Fuzzy c-means can also be also used as an alternative to estimate $\Bar{\psi}(\tau)$. The modified EC weights, hereafter termed as $w_{\mathrm{EC-2}}(k,\tau)$ are then estimated as the product of $\Bar{\psi}(\tau)$ and $\Bar{\lambda}(k,\tau)$ as
\begin{equation}
    w_{\mathrm{EC-2}}(k,\tau) = \Bar{\psi}(\tau) \times \Bar{\lambda}(k,\tau).
\end{equation}

\subsection{Improved EC-3}
In both improved EC-1 and EC-2, we used the average DOA unit vector to identify time-frames dominated by a single source using ${\psi}(\tau)$ and $\Bar{\psi}(\tau)$, respectively. Moreover, in improved EC-2 we employed additional clustering step to provide more weighting to the TF points dominated by a single source. As such, the computational complexity of EC-2 is higher than EC-1. As an alternative, in the improved EC-3 scheme, we use the mean of $\Bar{\lambda}(k,\tau)$ weights in a single time-frame to replace ${\psi}(\tau)$ and $\Bar{\psi}(\tau)$ with $\Hat{\psi}(\tau)$ as
\begin{equation}
    \Hat{\psi}(\tau) = \dfrac{1}{K}\sum_{i=1}^{K}\Bar{\lambda}(k_i,\tau).
\end{equation}

\noindent The use of $\Bar{\lambda}(k,\tau)$ values to identify single source-dominant time-frame is more robust under reverberation. The improved EC weights, hereafter termed as $w_{\mathrm{EC-3}}(k,\tau)$ are then estimated as the product of $ \Hat{\psi}(\tau)$ and $\Bar{\lambda}(k,\tau)$ as
\begin{equation}
    w_{\mathrm{EC-3}}(k,\tau) = \Hat{\psi}(\tau) \times \Bar{\lambda}(k,\tau).
\end{equation}

\par It should be noted that for all the above proposed schemes, we subsequently perform subsampling to select $P\%$ DOA unit vectors with highest weights. After clustering these subsampled DOA unit vectors, we cluster them to estimate the cluster centroids which are then associated to the source DOA.

\section{Simulations \& Results}
\label{sec:sim_results}

\begin{figure}[t!]
  \includegraphics[width=\columnwidth]{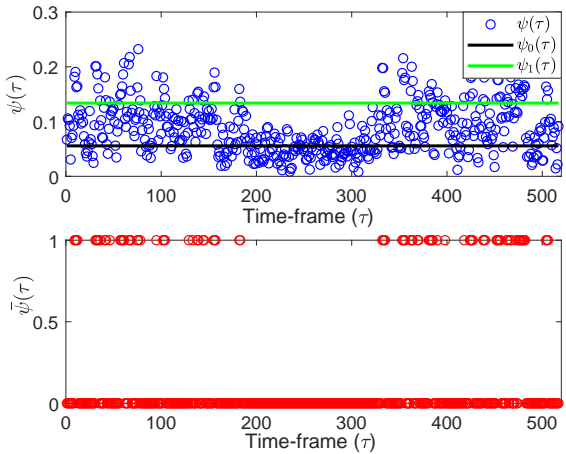}
  \caption{Estimation of $\Bar{\psi}(\tau)$ after the clustering of $\psi(\tau)$. The cluster centroids $\psi_0(\tau)$ and $\psi_1(\tau)$ are also depicted.}
  \label{fig:kmeans}
\end{figure}

The proposed algorithm, consisting of the multiple weighting schemes are evaluated using simulated data and the performance is compared with the EC method proposed in~\cite{hafezi2017multiple}. To generate the simulated 32-element rigid SMA recordings, we used the Spherical Microphone arrays Impulse Response Generator (SMIR-gen)~\cite{smirgen}. The SMA with a radius of $4.2~\mathrm{cm} $ is placed at the center of a room with dimensions $5 \times 6  \times 4~\mathrm{m} $ and the virtual acoustic sources are placed around the room. For each trial, the sources are randomly placed on a circle of radius $1$~m centered at the SMA position. For source signals, four speech utterances sampled at $16$ kHz are randomly selected from the TIMIT database~\cite{timit}. The generated impulse responses are then convolved with these speech utterances and then white Gaussian noise is added to these convolved signals such that the signal-to-noise ratio (SNR) is fixed at 20 dB. For the STFT analysis, FFT size of $1024$ with overlap $75\%$ is used. For the subset DOA selection, we also set $P = 5\%$ as in~\cite{hafezi2017multiple}. The accuracy of the DOA estimation is evaluated using the angular error between the unit vector corresponding to the estimated DOA, $\mathbf{u}_{\mathrm{est}}$ and the unit vector corresponding to the ground truth DOA $\mathbf{u}_{\mathrm{true}}$. This angular error $ \epsilon_{\mathbf{u}_{\mathrm{true}},\mathbf{u}_{\mathrm{est}}}$ can be mathematically expressed as  
\begin{equation}
    \epsilon_{\mathbf{u}_{\mathrm{true}},\mathbf{u}_{\mathrm{est}}} = \cos^{-1}(\mathbf{u}_{\mathrm{true}}^{T}\mathbf{u}_{\mathrm{est}})
\end{equation}

To evaluate the performance of the algorithms, we compute the angular error when the source spacing is varied from $5^\circ$ to $90^\circ$. For each spacing, we conduct $100$ trials with the sources placed around the room, while maintaining the source spacing. For each trial, the DOA error for each source is averaged.
Figure~\ref{fig:performance} shows the median of the angular error obtained from $100$ such trials using the proposed schemes as well as the EC approach for various source spacing. Compared to EC weight estimation~\cite{hafezi2017multiple}, the proposed improved EC-1, which uses  $\Bar{\lambda}(k,\tau)$ instead of ${\lambda}(k,\tau)$ achieves better performance for less source spacing. The improved EC-2 achieves the better performance than EC-1. This can be attributed to the clustering performed on $\psi(\tau)$ and using $\Bar{\psi}(\tau)$ for boosting the EC weights. However, the best performance is achieved by improved EC-3 approach which uses $\Hat{\psi}(\tau)$ and $ \Bar{\lambda}(k,\tau)$. 

\section{Conclusions}

\label{conclusion}
We proposed three schemes for improving the DOA estimation using estimation consistency (EC) approach. We highlighted the limitations of the EC approach under the effects of reverberation and noise. To address these limitations, we proposed the identification and usage of a DOA unit vector from the cluster of accurate DOA unit vectors using L2-distance criterion. We performed simulations using a spherical microphone array to compare and evaluate the improved EC weighting schemes along with the baseline EC approach. Our results demonstrate that the proposed schemes achieve better performance in terms of localization accuracy, especially when the acoustic sources are spatially close to each other. 
\begin{figure}[t!]
  \includegraphics[width=\columnwidth]{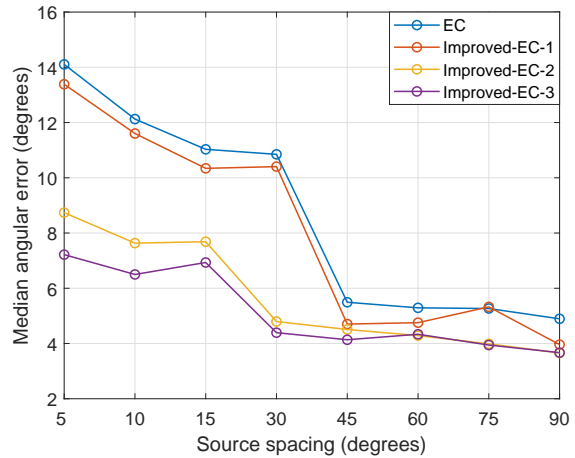}
  \caption{Median angular error using the proposed schemes as well as the EC approach for various source spacing.}
  \label{fig:performance}
\end{figure}

\bibliographystyle{IEEEtran}
%

\end{sloppy}
\end{document}